From carmelo@uevora.pt Mon Jan  6 11:07:23 1997
Date: Mon, 6 Jan 1997 10:50:50 GMT
From: Jose Manuel Carmelo <carmelo@uevora.pt>
To: peres@uevora.pt
Subject: file

				    22 December 1996

			 Dear Sirs,

  We are inclosing the manuscript LV5792 which we are resubmiting
for publication in Physical Review Letters. We are sending in
separate messages the LaTEX files for the letter to the Editor
and reference [4] of our manuscript.

		       Yours sincerely,

			 Jose Carmelo

\documentstyle[aps,version2,preprint]{revtex}    
\begin{document}

\draft

\begin{title}
Complete Pseudoparticle Representation and the Frequency-Dependent\\
Conductivity of the Hubbard chain           
\end{title}

\author{J. M. P. Carmelo$^{1}$, N. M. R. Peres $^{1,2}$,
and D. K. Campbell$^{2}$}
\begin{instit}
$^{1}$ Department of Physics, University of \'Evora,
Apartado 94, P-7001 \'Evora Codex, Portugal
\end{instit}
\begin{instit}
$^{2}$ Department of Physics, UIUC, 1110 West Green Street,
Urbana, IL 61801-3080
\end{instit}
\receipt{12 August 1996}

\begin{abstract}
We use a complete pseudoparticle operator representation to 
study the explicit form of the finite-frequency conductivity 
for the Hubbard chain. Our study reveals that the spectral 
weight is mostly concentrated at the $\omega =0$ Drude peak 
(except at density $n=1$) and at an absorption starting 
just below twice the value of the chemical potential
whose $\omega $-dependence we evaluate exactly. We also 
obtain the exact $\omega $ dependence of the higher-energy,
less-pronounced absorption edges, whose weights vanish for large on-site
interaction $U$.
\end{abstract}
\renewcommand{\baselinestretch}{1.656}   

\pacs{PACS numbers: 72.15. Nj, 05.30. Fk, 72.90.+y, 03.65. Ca}

\narrowtext

The unusual spectral properties of electronic Luttinger liquids 
\cite{Haldane1} imply that they cannot be described by 
one-electron, Fermi-liquid-like models. Further, descriptions
of these quantum systems in terms of exotic excitations such as 
holons and spinons, or pseudoparticles have been limited to the 
low-energy Hilbert subspace, and both bosonization and 
standard conformal-field theory techniques also apply only in 
that limit. On the other hand, the unusual  
finite-frequency transport properties observed in low-dimensional 
materials are far of being well understood. 
Thus developing a microscopic operator description of the 
finite-frequency/energy physics of low-dimensional electronic quantum 
liquids remains an important chalenge.

In this Letter we show that, for the {\it integrable} 
Luttinger liquid associated with the Hubbard chain 
\cite{Lieb,Takahashi} in a magnetic field, the generalized 
pseudoparticle operator representation introduced recently in 
Ref. \cite{Carmelo96c} provides the desired microscopic description 
for the finite-frequency/energy transport quantities. We find that the 
excitations that dominate the frequency-dependent conductivity at finite 
energies involve a small density of excited pseudoparticles.
Using the generalized pseudoparticle algebra we obtain,
in addition for the Drude peak (for the non-half-filled case),
explicit expressions for a prominent feature that sets in with a 
power law at $\omega \simeq 2 |\mu|-\delta$, where $\mu$ is the 
chemical potential and $\delta<<|\mu|$ is a small positive
energy we define below (which vanishes at large $U$ and at half 
filling for all values of $U$). We also obtain expressions for 
the less-pronounced, higher-energy absorptions of the conductivity spectrum. 
The fact that at {\it all} energy scales there is only zero-momentum 
pseudoparticle forward scattering in the Hubbard chain
confirms the recent indications that the integrability of BA 
solvable many-electron problems is closely related to peculiar
spectral and transport properties \cite{Prelo}. 
Although this exact property holds only for {\it integrable}
systems, we believe that our pseudoparticle operator description 
will provide the most suitable starting point for description of 
the unusual finite-frequency transport properties of both 
integrable and non-integrable electronic Luttinger liquids.

We consider the Hubbard chain \cite{Lieb,Takahashi} in a 
magnetic field $H$ and chemical potential $\mu$ 
\cite{Lieb,Takahashi,Carmelo92,Carmelo94,Carmelo96}, 
$\hat{H} = \hat{H}_{SO(4)} + 2\mu{\hat{S }}^c_z + 
2\mu_0 H{\hat{S}}^s_z$, where 

\begin{equation}
\hat{H}_{SO(4)} = -t\sum_{j,\sigma}
\left[c_{j\sigma}^{\dag }c_{j+1\sigma} +
c_{j+1\sigma}^{\dag }c_{j\sigma}\right] +
U\sum_{j} [c_{j\uparrow}^{\dag }c_{j\uparrow}-{1\over 2}]
[c_{j\downarrow}^{\dag }c_{j\downarrow}-{1\over 2}] \, .
\end{equation}
The diagonal generators of the $\eta$-spin and spin $SU(2)$ algebra are 
\cite{Carmelo96c} ${\hat{S}}^c_z = -{1\over 2}[N_a - \sum_{\sigma}
\hat{N}_{\sigma }]$ and ${\hat{S}}^s_z = -{1\over 2}\sum_{\sigma}
\sigma\hat{N}_{\sigma }$, respectively. 
Here we introduced the notations $\eta=S^c$ 
and $S=S^s$ for $\eta$-spin and spin, respectively,
$c_{j\sigma}^{\dagger}$ and $c_{j\sigma}$ are electron operators 
of spin projection $\sigma $ at site $j$, $\hat{N}_{\sigma }=\sum_{j} 
c_{j\sigma }^{\dagger }c_{j\sigma }$, and $N_a$, $t$, $U$, $\mu$, $H$, and 
$\mu _0$ are the number of sites, the transfer integral, the on-site 
Coulomb interaction, the chemical potential, the magnetic field, 
and the Bohr magneton, respectively. We consider the $\sigma $
electronic density $n_{\sigma}={N_{\sigma}\over N_a}$,
the electronic density $n=n_{\uparrow }+n_{\downarrow}$, the spin 
density $m=n_{\uparrow}-n_{\downarrow}$, and the $\sigma$
Fermi momentum $k_{F\sigma}=\pi n_{\sigma}$.
For simplicity we restrict our study to densities
$0<n\leq 1$ and spin densities $0<m\leq n$.

The recently introduced generalized pseudoparticle operator
representation refers to the whole Hilbert space of the Hubbard 
chain \cite{Carmelo96c} and involves the $\alpha,\gamma$ 
pseudoparticles with quantum numbers $\alpha $ and $\gamma $ such that
$\alpha =c,s$ and $\gamma =0,1,2,...,\infty$. The 
heavy pseudoparticles have $\gamma >0$ (whose bands have
energy gaps) whereas the light $\alpha,0$ pseudoparticles
are nothing but the usual $\alpha $ pseudoparticles of
the low-energy pseudoparticle theory 
\cite{Carmelo92,Carmelo94}. Following Refs.
\cite{Carmelo96c,Carmelo96}, as an alternative
to the $\alpha,0$ pseudoparticles, one can refer to their
holes, which are called pseudoholes. The use of the
pseudohole representation is required for the description 
in the present operator basis of the six generators of the 
$\eta $-spin and spin algebras. However, since the current operator 
associated with the conductivity spectrum commutes with all these 
generators, it projects the ground state (GS) onto a Hilbert 
subspace spanned by the lowest-weight states (LWS's) of these 
algebras. In that space $S^{\alpha }_z=
-S^{\alpha}$ and we can omit the pseudohole representation 
and express all quantities in terms of $\alpha,\gamma $ pseudoparticles 
(with $\gamma =0,1,2,...$) only. 

There are an infinite number of conservation laws associated with
the pseudoparticle numbers $N_{\alpha,\gamma}$, each Hamiltonian 
eigenstate having fixed values for these numbers. 
The $\alpha,\gamma$ pseudoparticles have creation (annihilation)
operator $b^{\dag }_{q,\alpha,\gamma }$ ($b_{q,\alpha,\gamma }$)
and obey an anticommuting algebra.
The pseudomomentum $q$ discrete values read 
$q_j={2\pi\over N_a}I^{\alpha,\gamma}_j$
where the quantum numbers $I^{\alpha,\gamma}_j$ can be
integers or half-odd integers depending on the parities of the 
pseudoparticle numbers. The excitations
associated with the changes in the integer or half-odd-integer
character of these numbers are generated by
the topological-momentum-shift operators whose expressions 
in terms of pseudoparticle operators are given in
Refs. \cite{Carmelo96c,Carmelo96}. These play an important 
role in the GS transitions we discuss below.

In our Hilbert subspace we have that
$S^{c }_z=-{1\over 2}[N_a - N_{c,0}
-\sum_{\gamma =1}^{\infty}2\gamma N_{c,\gamma}]$,
$S^{s }_z=-{1\over 2}[N_{c,0} - 2N_{s,0}
-\sum_{\gamma =1}^{\infty}2(1 +\gamma) N_{s,\gamma}]$,
and the total-momentum eigenvalue is $P=\sum_{q,\alpha}
\sum_{\gamma =0}^{\infty} qC_{\alpha,\gamma}
N_{\alpha,\gamma}(q)+\pi\sum_{\gamma =1}
(1+\gamma )N_{c,\gamma}$, where $C_{c,\gamma}=-1$ for 
$\gamma >0$ and $C_{\alpha,\gamma}=1$ otherwise. 
The pseudoparticles have bands $\epsilon^0_{\alpha,\gamma}(q)$ 
and group velocities $v_{\alpha ,\gamma}(q)={d\epsilon^0_{\alpha ,\gamma}(q) 
\over {dq}}$ and interact only via zero-momentum foraward scattering
associated with the pseudoparticle-Hamiltonian $f$ functions.
The $f$-function expression involves the
phase shifts $\Phi_{\alpha,\gamma;\alpha',\gamma'}
(q,q')$ and the pseudoparticle velocities. (The expressions of 
these quantities are given in Ref. \cite{Carmelo96c}.) The 
``light'' velocities $v_{\alpha ,\gamma}\equiv 
v_{\alpha ,\gamma}(q_{F\alpha ,\gamma})$, where the pseudo-Fermi 
momentum is $q_{F\alpha ,\gamma}={\pi N_{\alpha,\gamma}\over
N_a}$, and the parameters

\begin{equation}
\xi^j_{\alpha ,\gamma ;\alpha',\gamma'}=\delta_{\alpha,\alpha'}
\delta_{\gamma,\gamma'} + \sum_{l=\pm 1}l^j
\Phi_{\alpha,\gamma;\alpha',\gamma'}(q_{F\alpha,\gamma},
lq_{F\alpha',\gamma'}) \, ,
\end{equation}
with $j=0,1$ play a central role in the quantum-liquid physics. 
The parameters $(2)$ are simple combinations of the 
two-pseudoparticle forward-scattering phase shifts 
\cite{Carmelo96c}. The relevant
excitations which contribute to the conductivity involve
a small density of pseudoparticles. For these excitations
the ratios ${\Delta N_{\alpha,0}\over N_a}$  
and ${N_{\alpha,\gamma}\over N_a}$ for $\gamma >0$ are
small, and the parameters $(2)$ are such that $\xi^j_{\alpha,0;\alpha',0} = 
\xi^j_{\alpha\alpha'}$, $\xi^j_{\alpha,\gamma;\alpha',\gamma'} = 
\delta_{\alpha,\alpha'}\delta_{\gamma,\gamma'}$, 
and $\xi^1_{\alpha,0;\alpha',\gamma'} = 
\xi^0_{\alpha,\gamma;\alpha',0} = 0$. 
Here $\gamma ,\gamma'>0$ and $\xi^j_{\alpha\alpha'}$ are
the low-energy parameters defined in the second paper
of Ref. \cite{Carmelo92}. Moreover, $\xi^0_{\alpha,0;\alpha',\gamma'} = 
2\Phi_{\alpha,0;\alpha',\gamma'}(q_{F\alpha,0},0)$  
and $\xi^1_{\alpha,\gamma;\alpha',0} =
2\Phi_{\alpha,\gamma;\alpha',0}(0,q_{F\alpha',0})$.  

At finite magnetization and away of half filling, all 
low-energy Hamiltonian eigenstates correspond to empty 
$\alpha,\gamma $ bands for $\gamma >0$. 
Each canonical ensemble (with constant $N_{\uparrow}$ and
$N_{\downarrow}$ electron numbers) contains different {\it 
sub-canonical ensembles} characterized by different sets of 
constant $N_{\alpha,\gamma}$ pseudoparticle numbers. 
In our parameter space these numbers obey the sum rules 
$N=N_{\uparrow}+N_{\downarrow} = N_{c,0}+\sum_{\gamma 
=1}^{\infty}2\gamma N_{c,\gamma}$ and $N_{\uparrow}-
N_{\downarrow} = N_{c,0} - 2N_{s,0} - \sum_{\gamma =1}^{\infty}
2(1+\gamma )N_{s,\gamma}$. The Hamiltonian(s) eigenstate(s) of 
lowest energy in each Hilbert subspace with constant $N_{\alpha,\gamma}$
numbers was called in Ref. \cite{Carmelo96c} generalized ground state 
(GGS). This is always a $\alpha,\gamma $ pseudoparticle filled pseudo-Fermi 
sea. [The ground state (GS) is a particular case of GGS where 
$N_{\alpha,\gamma}=0$ for $\gamma >0$.] 

A remarkable property is that the transition between
a GS and {\it any} Hamiltonian eigenstate can be
separated into two types of excitations: (a) a topological 
GS -- GGS transition which involves the creation or annihilation
of pseudoparticles as well as the
occurrence of topological-momentum shifts and (b) a {\it 
Landau-liquid} excitation \cite{Carmelo94,Carmelo96}
associated with pseudoparticle - pseudohole processes relative to 
the GGS. In general the topological transitions (a) 
are basically superpositions of three kinds of elementary 
transitions: (i) GS - GS transitions involving changes in
the $\sigma $ electron numbers by one, (ii) transitions
which lead to non-LWS's and non-HWS's outside the BA and
are described by the pseudoholes of Refs. \cite{Carmelo96c,Carmelo96}
-- these transitions cannot be induced by the current operator
and lead to states outside our Hilbert subspace --
and (iii) creation of single $\alpha,\gamma$ pseudoparticles at 
constant values of $\eta$, $\eta_z$, $S$, and $S_z$. 
While the transitions (i) are gapless, the elementary
excitations (ii) and (iii) require a finite amount of 
energy. In the case of the latter excitations 
the gaps, $\omega_0 =E^0_{GGS}-E^0_{GS}$, 
where $E^0_{GGS}$ and $E^0_{GS}$ are the GGS and GS
energy, respectively, are for small densities
${N_{\alpha,\gamma}\over N_a}$ (with $\gamma >0$)
given by \cite{Carmelo96c}

\begin{equation}
\omega_0 = 2|\mu|\sum_{\gamma =1}\gamma N_{c,\gamma} 
+ 2\mu_0|H|\sum_{\gamma =1}(1+\gamma ) 
N_{s,\gamma} + \sum_{\alpha,\gamma =1}\epsilon^0_{\alpha ,\gamma }(0)
N_{\alpha,\gamma} \, .
\end{equation}
The generators of the excitations (i) were 
studied in Ref. \cite{Carmelo96} and the ones of (ii) involve
only topological-momentum shifts and pseudohole processes 
which describe either creation of electron pairs or spin flip 
processes. The important transitions for the conductivity are 
of type (iii). Consider as an example of such excitation
the GS - GGS transition corresponding to the creation of 
one zero-momentum $c,\gamma $ pseudoparticle which we find below to be the
relevant elementary process for the finite-frequency conductivity. 
The zero-momentum condition requires $\gamma $ to be odd.
Such transition involves one $s\equiv s,0$ topological-momentum 
shift of momentum $k=\pm k_{F\downarrow}$, a zero-momentum
excitation involving the annihilation of a number $2\gamma$ 
of $c,0$ pseudoparticles, a $k=\mp k_{F\downarrow}$ excitation
associated with the annihilation of a number $\gamma$ of $s,0$ 
pseudoparticles, and the creation of one $c,\gamma$ pseudoparticle at
$q=0$. 

The infinity of pseudoparticle-number
conservation laws implies that, for Lorentz invariance,
each $\alpha ,\gamma$ channel is an independent problem. Therefore, the
study of the general pseudoparticle-energy spectrum of 
Ref. \cite{Carmelo96c} as $k,(\omega -\omega_0)\rightarrow 0$
reveals that the problem becomes critical, leading to
a small momentum $k$ and low-energy $(\omega -\omega_0)$
conformal-field theory \cite{Carmelo96d}. Furthermore,
this reveals that the relevant GS transitions which dominate the
charge -charge (and spin - spin) correlation functions at
$k=0$ occur at low-energy $(\omega -\omega_0)$, with the
possible $\omega_0$ values given in Eq. $(3)$.

We can write the charge-current operator 
as ${\hat{J}}={\hat{J}}_d+{\hat{J}}_{off}$, where
${\hat{J}}_d=\sum_{l}\langle l|{\hat{J}}|l\rangle
|l\rangle\langle l|$ and ${\hat{J}}_{off}=\sum_{l\neq l'}\langle l|
{\hat{J}}|l'\rangle |l\rangle\langle l'|$, and the summations run
over all Hamiltonian eigenstates. By applying a vector potential,
we can derive the expression for ${\hat{J}}_d$.
As at low energy \cite{Carmelo92}, the pseudoparticles are the 
transport carriers at all energy scales and the electronic degrees 
of freedom couple to the potential through them. We find that the 
couplings of the $\alpha,\gamma$ pseudoparticles to the potential 
read $-e{\cal C}^{\rho}_{\alpha,\gamma}$ where $-e$ is the electronic 
charge, the superscript $\rho$ indicates the coupling to charge, and 
${\cal C}^{\rho}_{c,\gamma}=\delta_{\gamma ,0}+2\gamma$ and 
${\cal C}^{\rho}_{s,\gamma}=0$. Like the normal-ordered 
pseudoparticle Hamiltonian introduced in Ref. \cite{Carmelo96c}, the 
GS-normal-ordered expression for ${\hat{J}}_d$ has an infinite number 
of terms, $:{\hat{J}}_d:=\sum_{i=1}^{\infty}
{\hat{J}}_d^{(i)}$. In this case only the first-order
term is relevant at low positive energy $(\omega -\omega_0)$.
It reads 

\begin{equation}
{\hat{J}}_d^{(1)}=-e\sum_{q,\alpha}\sum_{\gamma 
=0}^{\infty}j_{\alpha,\gamma}(q):{\hat{N}}_{\alpha,\gamma}(q): \, ,
\end{equation}
where the pseudoparticle elementary charge currents are
given by 

\begin{equation}
j_{\alpha,\gamma}(q) = \sum_{\alpha'}\sum_{\gamma'=0}^{\infty} 
{\cal C}^{\rho}_{\alpha',\gamma'}\left[
v_{\alpha,\gamma}(q)\delta_{\alpha ,\alpha'}
\delta_{\gamma ,\gamma'} + {1\over 2\pi}
\sum_{j=\pm 1}jf_{\alpha,\gamma;\alpha'
,\gamma'}(q,jq_{F\alpha',\gamma'})\right] \, ,
\end{equation}
and involve only the group velocities and $f$ functions.
The pseudoparticle current $j_{\alpha,\gamma}\equiv 
j_{\alpha,\gamma}(q_{F\alpha,\gamma})$ and charge transport 
mass 

\begin{equation}
m^{\rho}_{\alpha,\gamma}\equiv {q_{F\alpha,\gamma}\over
{\cal C}^{\rho}_{\alpha,\gamma}j_{\alpha,\gamma}} \, ,
\end{equation}
such that $m^{\rho}_{c,\gamma}=q_{F\alpha,\gamma}/
[(\delta_{\gamma ,0}+2\gamma)j_{\alpha,\gamma}]$
and $m^{\rho}_{s,\gamma}=\infty$, play central roles 
in charge transport. (The $\gamma =0$ expressions
recover the results of \cite{Carmelo92}.) 

The frequency-dependent conductivity, Re$\sigma (\omega)$,  
involves transitions induced by ${\hat{J}}_{off}$. For the 
Hubbard model at intermediate and large values of $U$ (in any 
dimension), it consists in principle of three 
different pieces: (1) the Drude absorption peak (absent for $n=1$), 
(2) a finite-frequency absorption associated with creation of new
doubly-occupied sites, and (3) an absorption continuum between the 
Drude peak and absorption (2) (absent for $n=1$). However, numerical 
studies show that (in contrast to 2D) the 1D Hubbard model has only an 
extremely small type-(3) absorption \cite{Horsch}. Since the conductivity 
sum rule \cite{Carmelo86,Zotos} proves that there is spectral weight 
away of the Drude peak, it must be mostly concentrated 
in the absorption (2). This picture is confirmed by our exact
study. Although we have not derived a closed-form expression
for ${\hat{J}}_{off}$, our operator basis does provide important 
information about the conductivity spectrum: namely, 1) The occurrence
of only forward scattering in the Landau-liquid term of
the pseudoparticle Hamiltonian is consistent with the
extremely small type-(3) absorption found numerically 
\cite{Horsch,Carmelo86,Zotos} and indicates that the
topological finite-energy transitions (a)
[followed by excitations (b)] 
are relevant for that spectrum at finite energy; 2) That the
current commutes with ${\bf {\eta}}^2$, ${\bf S}^2$, 
${\hat{\eta}}_z$, and ${\hat{S}}_z$ restricts the relevant topological
excitations (a) to superpositions of elementary transitions
(iii); 3) That the $s,\gamma$ pseudoparticles
do not couple to charge further limits the transitions (iii)
to the ones associated with creation of $c,\gamma$ 
pseudoparticles and the important contributions to the 
conductivity spectrum occur for small values of $\omega $ and 
for values of $\omega $ just above
the finite frequencies of general form 

\begin{equation}
\omega_0=\sum_{\gamma =1}^{\infty}\Bigl(2\gamma|\mu|
+ \epsilon^0_{c ,\gamma }(0)\Bigl)N_{c,\gamma} \, ;
\end{equation}
4) Since the conductivity spectrum corresponds to
$k=0$ momentum excitations, only GS - GGS transitions 
such that $\sum_{\gamma =1}^{\infty}(1+\gamma )N_{c,\gamma}$
is even are allowed; 5) Finally, based on the fully non-dissipative
character of the pseudoparticle interactions at all
energy scales combined with the occurrence of the
infinite pseudoparticle-number conservation laws
associated with the above small-$k$ and low-$(\omega -\omega_0)$
generalized conformal-field theory \cite{Carmelo96d},
we could evaluate the charge - charge correlation
function, $\chi (k,\omega)$, at small $k$ and $\omega $
just above the $\omega_0$ values $(7)$.
For small $\omega $ the charge - charge correlation function leads
to the Drude peak \cite{Carmelo92,Carmelo86,Zotos} only, 
$(-e)^22\pi[q_{Fc,0}/m^{\rho}_{c,0}]\delta (\omega)$. 
In order to evaluate the conductivity spectrum
${\rm Re}\sigma (\omega)$ we use its relation
to the charge-charge correlation function, 
${\rm Re}\sigma (\omega)\propto 
\lim_{k\to 0}{\omega\chi (k,\omega)\over k^2}$.
We have evaluated for low positive energies $(\omega -\omega_0)$
the following conductivity general expression for type-(2) absorption:

\begin{equation}
{\rm Re}\sigma (\omega) = C(\{N_{c,\gamma }\})
\Bigl(\omega -\omega_0\Bigl)^{\zeta } \, ,
\end{equation}
where $\omega_0$ is given by Eq. $(7)$, $\{N_{c,\gamma} \}$ stands 
for the set of numbers $N_{c,1},N_{c,2},...,N_{c,\infty}$,
and the critical exponent reads

\begin{equation}
\zeta = - 2 + 
{1\over 2}\sum_{\gamma =1}^{\infty}[N_{c,\gamma}]^2
+ {1\over 2}\sum_{\alpha}\Bigl(\sum_{\gamma =1}^{\infty}
[\gamma (2\xi^0_{\alpha c}+\xi^0_{\alpha s})
-\xi^0_{\alpha ,0;c,\gamma}]N_{c,\gamma}\Bigl)^2\, .
\end{equation}
The constant $C(\{N_{c,\gamma} \})$ is such that
$C(\{N_{c,\gamma} \})\rightarrow 0$ in the limits $U\rightarrow 0$,
$U\rightarrow\infty$ for $m=0$, and $m\rightarrow n$. This is 
because in these limits the Drude peak exausts the 
conductivity sum rule \cite{Carmelo86,Zotos}.
Moreover, at large $U>>4t$ and $m=0$, $C(\{N_{c,\gamma} \})\rightarrow 0$
except for $N_{c,1}=1$ and $N_{c,\gamma }=0$ when
$\gamma >1$, $C(N_{c,1}=1)$ remaining finite. This
follows from the fact that in that limit the GS normal-ordered double
occupation operator $:\hat{D}:=\hat{D}-\langle GS|\hat{D}|GS\rangle$,
where $\hat{D}=c_{j\uparrow}^{\dag }c_{j\uparrow}
c_{j\downarrow}^{\dag }c_{j\downarrow}$, reads in the
pseudoparticle basis

\begin{equation}
:\hat{D}:=\sum_{\gamma =1}^{\infty}\gamma{\hat{N}}_{c,\gamma} \, ,
\hspace{1cm} U>>4t \, .
\end{equation}
On the other hand, since in this limit the electric-current operator 
only induces GS transitions associated with doubly-occupancy changes 
$\Delta D=0$ and $\Delta D=1$, it follows that the only GS - GGS 
transition allowed is the creation of one $c,1$ pseudoparticle. 
It is this selection rule which justifies that at large $U$
the only finite-energy absorption in the frequency-dependent 
conductivity is generated by such transition followed by 
pseudoparticle -- pseudohole processes around the GGS.
(For large $U$ such elementary processes create one doubly-occupied 
site.) In addition, for arbitrary values of $U$ expression $(9)$
shows that for all allowed $c,\gamma $ GS - GGS transitions 
$\zeta >1$ {\it except} for the one-$c,1$-pseudoparticle
absorption edge, for which ${1\over 2}\leq\zeta <1$. 
For instance, for the transition
creating two $c,1$ pseudoparticles the exponent $(9)$ is
such that ${14\over 2}\leq\zeta <9$. Since the absorption 
associated with other energies and excitations is extremely 
small, we thus conclude that in the 1D 
model [in contrast to the
2D case where the absorption (3) is important \cite{Horsch}]
the spectral weight is mostly concentrated 
at the $\omega =0$ Drude peak (for $n\neq 1$) and at an absorption
starting at $\omega\sim 2|\mu |-\delta$, where
$\delta =-\epsilon^0_{c,1}(0)$. Here the 
conductivity shows a Luttinger-liquid band edge with infinite 
slope at $\omega =2|\mu |-\delta$. 
For fixed values of $U$ and $m$, $2|\mu |$ changes between 
$2|\mu |=\Delta_{MH}$ for $n\rightarrow 1$ and a maximum value
of $2|\mu |=\Delta_{MH}+8t$ for $n\rightarrow 0$, 
where $\Delta_{MH}$ is the half-filled Mott-Hubbard gap \cite{Lieb} 
(which is an increasing function of
$U$ and $m$). On the other hand, $\delta$ 
changes between $\delta=0$
for $n\rightarrow 1$ and a maximum value for $n\rightarrow 0$. 
Based on the form of the $c,1$ excitation spectrum we predict the
above absorption to have a width $W$ such that $0<W\leq 8t$. $W$ vanishes 
for $n\rightarrow 0$ and reaches its 
maximum value, $W=8t$, for $n\rightarrow 1$. This $n=1$ result
and the fact that for all densities and large $U$ only the 
one-$c,1$-pseudoparticle absorption is present in the conductivity 
spectrum at finite frequencies, fully agrees with the numerical 
studies of Refs. \cite{Fye,Campbell}. [For smaller values of $U$, 
considerably smaller weight exists at 
the remaining absorptions of higher energy and vanishing slope;
the slope also vanishes for the extremely small absorption (3), which
for the doped Mott-Hubbard insulator and $\omega\rightarrow 0$, is 
predicted to have an $\omega^{3\over 2}$ onset \cite{Horsch}.]
The one-$c,1$-pseudoparticle absorption edge exponent approaches 
its minimal value, $\zeta ={1\over 2}$, when the
amount of spectral weight away of the Drude peak is largest
(at zero field and finite $U$, $\zeta={1\over 2}$ for
half filling) and $\zeta\rightarrow 1$ in the limits 
$U\rightarrow\infty$ (at $m=0$) and $m\rightarrow n$ 
when all spectral weight is transferred to the Drude peak [and the 
absorption (2) disappears]. It follows from the above analysis
that at half filling, $m=0$, finite values of $U$,
and small values of $\omega -\Delta_{MH}$
the $c,1$ transitions lead to an absorption edge,
${\rm Re}\sigma (\omega) = C(N_{c,1}=1)
\sqrt{\omega -\Delta_{MH}}$. (Due to the absence of Drude peak 
at $n=1$, the corresponding absorption contains most of the
conductivity spectral weight.)

We thank J.M.B. Lopes dos Santos and J. Tinka Gammel for illuminating 
discussions and the US NSF (Grant DMR-8920538) for partial support.
N.M.R.P. thanks the Luso-American and Gulbenkian Foundations
for financial support.


\end{document}